# FINITE ELEMENT MODEL UPDATING USING RESPONSE SURFACE METHOD

Tshilidzi Marwala
School of Electrical and Information Engineering
University of the Witwatersrand
P/Bag 3, Wits, 2050, South Africa
t.marwala@ee.wits.ac.za

This paper proposes the response surface method for finite element model updating. The response surface method is implemented by approximating the finite element model surface response equation by a multi-layer perceptron. The updated parameters of the finite element model were calculated using genetic algorithm by optimizing the surface response equation. The proposed method was compared to the existing methods that use simulated annealing or genetic algorithm together with a full finite element model for finite element model updating. The proposed method was tested on an unsymmetrical H-shaped structure. It was observed that the proposed method gave the updated natural frequencies and mode shapes that were of the same order of accuracy as those given by simulated annealing and genetic algorithm. Furthermore, it was observed that the response surface method achieved these results at a computational speed that was more than 2.5 times as fast as the genetic algorithm and a full finite element model and 24 times faster than the simulated annealing.

## Introduction

Finite element (FE) models are widely used to predict the dynamic characteristics of aerospace structures. These models often give results that differ from the measured results and therefore need to be updated to match the measured data. FE model updating entails tuning the model so that it can better reflect the measured data from the physical structure being modeled[1]. One fundamental characteristic of an FE model is that it can never be a true reflection of the physical structure but it will forever be an approximation. FE model updating fundamentally implies that we are identifying a better approximation model of the physical structure than the original model. The aim of this paper is to introduce updating of finite element models using Response Surface Method (RSM)[2]. Thus far, the RSM method has not been used to solve the FE updating problem[1]. This new approach to FE model updating is compared to methods that use simulated annealing (SA) or genetic algorithm (GA) together with full FE models for FE model updating. FE model updating methods have been implemented using different types of optimization methods such as genetic algorithm and conjugate gradient methods[3-5]. Levin and Lieven[5] proposed the use of SA and GA for FE updating.

RSM is an approximate optimization method that looks at various design variables and their responses and identify the combination of design variables that give the best response. The best response, in this paper, is defined as the one that gives the minimum distance between the measured data and the data predicted by the FE model. RSM attempts to replace implicit functions of the original design optimization problem with an approximation model, which traditionally is a polynomial and therefore is less expensive to evaluate. This makes RSM very useful to FE model updating because optimizing the FE model to match measured data to FE model generated data is a computationally expensive exercise. Furthermore, the calculation of the gradients that are essential when traditional optimization methods, such as conjugate gradient methods, are used is computationally expensive and often encounters numerical problems such as ill-conditioning. RSM tends to be immune to such problems when used for FE model updating. This is largely because RSM solves a crude approximation of the FE model rather than the full FE model which is of high dimensional order. The multi-layer perceptron (MLP)[6] is used to approximate the response equation. The RSM is particularly useful for optimizing systems that are evolving as a function of time, a situation that is prevalent in model-based fault diagnostics found in the manufacturing sector. To date, RSM has been used extensively to optimize complex models and processes[7,8].

In summary, the RSM is used because of the following reasons: (1) the ease of implementation that includes low computational time; (2) the suitability of the approach to the manufacturing sector where model-based methods are often used to monitor structures that evolve as a function of time.

FE model updating has been used widely to detect damage in structures[9]. When implementing FE updating methods for damage identification, it is assumed that the FE model is a true dynamic representation of the structure and this is achieved through FE model updating. This means that changing any physical parameter of an element in the FE model is equivalent to introducing damage in that region. There are two approaches that are used in FE updating: direct methods and iterative methods[1]. Direct methods, which use the modal properties, are computationally efficient to implement and reproduce the measured modal data exactly. Furthermore, they do not take into account the physical parameters that are updated. Consequently, even though the FE model is able to predict measured quantities, the updated model is limited in the following ways: it may lack the connectivity of nodes - connectivity of nodes is a phenomenon that occurs naturally in finite element modeling because of the physical reality that the structure is connected; the updated matrices are populated instead of banded - the fact that structural elements are only connected to their neighbors ensures that the mass and stiffness matrices are diagonally dominated with few couplings between elements that are far apart; and there is a possible loss of symmetry of the systems matrices. Iterative procedures use changes in physical parameters to update FE models and produce models that are physically realistic. Iterative methods that use modal properties and the RSM for FE model updating are implemented in this paper. The FE models are updated so that the measured modal properties match the FE model predicted modal properties. The proposed RSM updating method is tested on an unsymmetrical H-shaped structure.

---

*Associate Professor

## Mathematical Background

In this study, modal properties, i.e. natural frequencies and mode shapes, are used as a basis for FE model updating. For this reason these parameters are described in this section. Modal properties are related to the physical properties of the structure. All elastic structures may be described in terms of their distributed mass, damping and stiffness matrices in the time domain through the following expression[10]:

$$[M]\{X''\} + [C]\{X'\} + [K]\{X\} = \{F\} \quad (1)$$

where *[M], [C]* and *[K]* are the mass, damping and stiffness matrices respectively, and *{X}, {X'}* and *{X''}* are the displacement, velocity and acceleration vectors respectively while *{F}* is the applied force vector. If equation 1 is transformed into the modal domain to form an eigenvalue equation for the $i^{th}$ mode, then[10]:

$$(-\overline{\omega}_i^2 [M] + j\overline{\omega}_i [C] + [K])\{\overline{\phi}\}_i = \{0\} \quad (2)$$

where $j = \sqrt{-1}$, $\overline{\omega}_i$ is the $i^{th}$ complex eigenvalue, with its imaginary part corresponding to the natural frequency $\omega_i$, *{0}* is the null vector and $\{\overline{\phi}\}_i$ is the $i^{th}$ complex mode shape vector with the real part corresponding to the normalized mode shape *{ϕ}$_i$*. From equation 2, it may be deduced that the changes in the mass and stiffness matrices cause changes in the modal properties of the structure. Therefore, the modal properties can be identified through the identification of the correct mass and stiffness matrices. The frequency response functions (FRFs) are defined as the ratio of the Fourier transformed response to the Fourier transformed force. The FRFs may be expressed in receptance and inertance form. On the one hand, receptance expression of the FRF is defined as the ratio of the Fourier transformed displacement to the Fourier transformed force. On the other hand, inertance expression of the FRF is defined as the ratio of the Fourier transformed acceleration to the Fourier transformed force. The inertance FRF *(H)* may be written in terms of the modal properties by using the modal summation equation as follows[10]:

$$H_{kl}(\omega) = \sum_{i=1}^{N} \frac{-\omega^2 \phi_k^i \phi_l^i}{-\omega^2 + 2\zeta_i \omega_i \omega j + \omega_i^2} \quad (3)$$

Equation 3 is an FRF due to excitation at position *k* and response measurement at position *l*, $\omega$ is the frequency point, $\omega_i$ is the $i^{th}$ natural frequency, *N* is the number of modes and $\zeta_i$ is the damping ratio of mode *i*. The excitation and response of the structure and Fourier transform method[10] can be used to calculate the FRFs. Through equation 3 and a technique called modal analysis[10], the natural frequencies and mode shapes can be indirectly calculated from the measured FRFs. The modal properties of a dynamic system depend on the mass and stiffness matrices of the system as indicated by equation 2. Therefore, the measured modal properties can be reproduced by the FE model if the correct mass and stiffness matrices are identified.

FE model updating is achieved by identifying the correct mass and stiffness matrices. The correct mass and stiffness matrices, in the light of the measured data, can be obtained by identifying the correct moduli of elasticity for various sections of the structure under consideration[1]. In this paper, to correctly identify the moduli of elasticity of the structure, the following cost function that measures the distance between measured data and FE model calculated data, is minimized:

$$E = \sum_{i=1}^{N} \gamma_i \left( \frac{\omega_i^m - \omega_i^{calc}}{\omega_i^m} \right)^2 \ldots$$
$$+ \beta \sum_i^N \left(1 - diag(MAC(\{\phi\}_i^{calc}, \{\phi\}_i^m))\right) \quad (4)$$

Here *m* is for measured, *calc* is for calculated, *N* is the number of modes; $\gamma_i$ is the weighting factor that measures the relative distance between the initial estimated natural frequencies for mode *i* and the target frequency of the same mode; the parameter $\beta$ is the weighting function on the mode shapes; the MAC is the modal assurance criterion[11]; and the $diag(MAC)_i$ stands for the $i^{th}$ diagonal element of the MAC matrix. The MAC is a measure of the correlation between two sets of mode shapes of the same dimension. In equation 4 the first part has a function of ensuring that the natural frequencies predicted by the FE model are as close to the measured ones as possible while the second term ensures that the mode shapes between measurements and those predicted by the FE model are correlated. When two sets of mode shapes are perfectly correlated then the MAC matrix is an identity matrix. The updated model is evaluated by comparing the natural frequencies and mode shapes from the FE models before and after updating to the measured ones.

## Response Surface Method

RSM method is a procedure that operates by generating a response for a given input. The inputs are the parameters to be updated and the response is the error between the measured data and the FE model generated data. Then an approximation model of the input parameters and the response, called a response surface equation, is constructed. As a consequence of this, the optimization method operates on the surface response. This equation is usually simple and not computationally intensive as opposed to a full FE model. RSM has other advantages such as the ease of implementation through parallel computation and the ease at which parameter sensitivity can be calculated.

The proposed RSM consists of these essential components: (1) the response surface approximation equation; and (2) the optimization procedure. There are many techniques that have been used for response surface approximation such as polynomial approximation[12] and neural networks[13]. A multi-layer perceptron is used as a response surface approximation equation[6]. Further understanding of different approaches to response surface approximation may be found in the literature[14-19]. In this paper, MLP is used because it has been successfully used to solve complicated regression problems. The details of the MLP are described in the next section. The second component of the RSM is the optimization of the response surface. There are many types of optimization methods that can be used to optimize the response surface equation and these include the gradient based methods[20] and evolutionary computation methods[21]. The gradient based methods have a shortcoming of identifying local optimum solutions while evolutionary computing methods are better able to identify global optimum solution. As a result of the global optimum advantage of evolutionary methods, in this study the GA is used to optimize the response surface equation. The manner in which the RSM is implemented is shown in Figure 1.

In this figure it shown that the RSM is implemented by following these steps:
1) Setting initial conditions which are: updating parameters, updating objective, which is in equation 4, and updating space.
2) The FE model is then used to generate sample response surface data
3) MLP is used to approximate the response surface approximation equation from the data generated in Step 2.
4) GA is used to find a global optimum solution.
5) The new optimum solution is used to evaluate the response from the full FE model.
6) If the optimum solution does not satisfy the objective, then the new optimum and the corresponding FE model calculated response replaces the candidate with the worst response in data set generated in Step 2 and then steps 3 to 5 are repeated. If the objective is satisfied then stop and the optimum solution becomes the ultimate solution.

Step 6 ensures that the simulation always operates in the region of the optimum solution. The next section

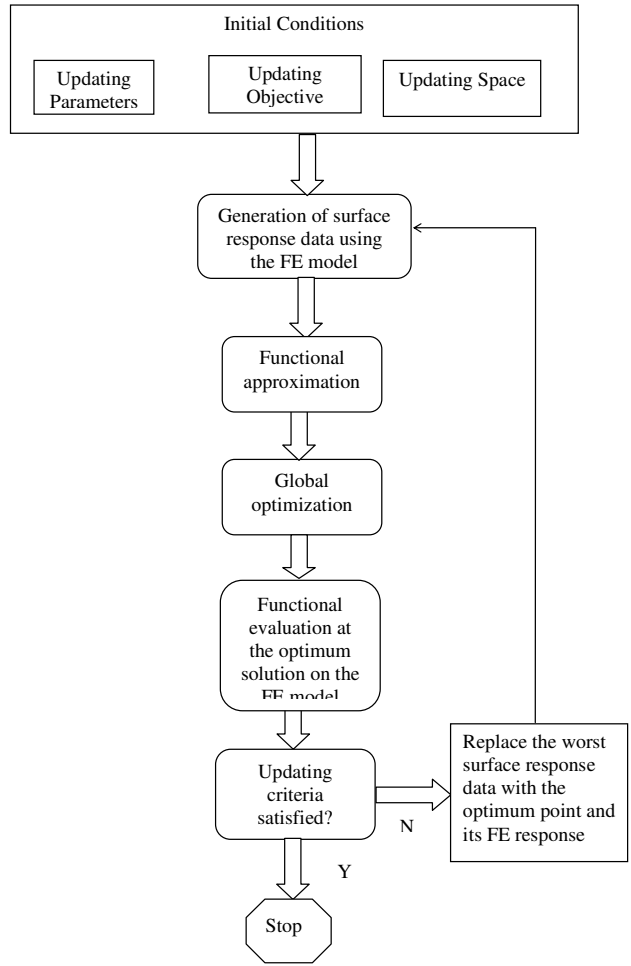

Figure 1. The flowchart of the RSM. Here N stands for no and Y stands for yes.

describes an MLP, which is used for functional approximation.

## Multi-layer Perceptron

Multi-layer perceptron is a type of neural networks which used in the present study. This section gives the over-view of the MLP in the context of functional approximation. The MLP is viewed in this paper as parameterized graphs that make probabilistic assumptions about data. Learning algorithms are viewed as methods for finding parameter values that look probable in the light of the data. Supervised learning is used to identify the mapping function between the updating parameters (*x*) and the response (*y*). The response is calculated using equation 4. The reason why the MLP is used is because it provides a distributed representation with respect to the input space due to cross-coupling between input, hidden and output layers. The MLP architecture contains a hyperbolic tangent basis function in the hidden units and linear basis functions in the output units[6]. A schematic illustration of the MLP is shown in Figure 2.

This network architecture contains hidden units and output units and has one hidden layer. The bias parameters in the first layer are shown as weights from an extra input having a fixed value of $x_0$=1. The bias parameters in the second layer are shown as weights from an extra hidden unit, with the activation fixed at $z_0$=1. The model in Figure 2 is able to take into account the intrinsic dimensionality of the data. Models of this form can approximate any continuous function to arbitrary accuracy if the number of hidden units *M* is sufficiently large. The relationship between the output *y*, representing error between the model and measured data, and input, *x*, representing updating parameters may be written as follows[6]:

$$y_k = \left( \sum_{j=1}^{M} w_{kj}^{(2)} \tanh\left( \sum_{i=1}^{d} w_{ji}^{(1)} x_i + w_{j0}^{(1)} \right) + w_{k0}^{(2)} \right) \quad (5)$$

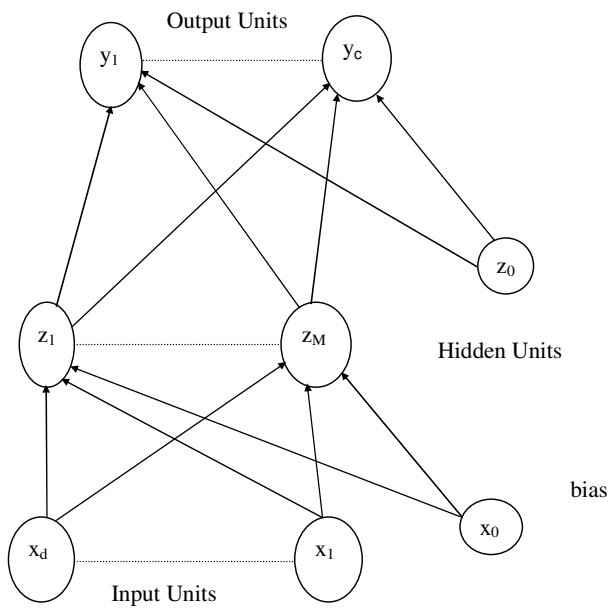

Figure 2. Feed-forward network having two layers of adaptive weights

Here, $w_{ji}^{(1)}$ and $w_{ji}^{(2)}$ indicate weights in the first and second layers, respectively, going from input *i* to hidden unit *j*, *M* is the number of input units, *d* is the number of output units while $w_{j0}^{(1)}$ indicates the bias for the hidden unit *j*. Training the neural network identifies the weights in equations 5 and a cost function must be chosen to identify these weights. A cost function is a mathematical representation of the overall objective of the problem. The main objective, this is used to construct a cost function, is to identify a set of neural network weights given updating parameters and the error between the FE model and the measured data. If the training set $D = \{x_k, t_k\}_{k=1}^{N}$ is used and assuming that the targets *t* are sampled independently given the inputs $x_k$ and the weight parameters, $w_{kj}$, the cost function, *E*, may be written as follows using the sum-of-square error function[6]:

$$E = \sum_{n=1}^{N} \sum_{k=1}^{K} \{t_{nk} - y_{nk}\}^2 \quad (6)$$

The sum-of-square error function is chosen because it has been found to be suited to regression problems[6]. In equation 6, *N* is the number of training examples and *K* is the number of output units. In this paper, *N* is equal to 150, while *K* is equal to 1.

Before the MLP is trained, the network architecture needs to be constructed by choosing the number of hidden units, *M*. If *M* is too small, the MLP will be insufficiently flexible and will give poor generalization of the data because of high bias. However, if *M* is too large, the neural network will be unnecessarily flexible and will give poor generalization due to a phenomenon known as over-fitting caused by high variance. In this study, we choose *M* such that the number of weights is at most fewer than the number of response data. This is in line with the basic mathematical principle which states that in order to solve a set of equations with *n* variables you need at least *n* independent data points. The next section describes the GA, which is a method that is used to solve for the optimum solution of the response surface approximation equation.

## Genetic Algorithms

GA was inspired by Darwin's theory of natural evolution. Genetic algorithm is a simulation of natural evolu-tion where the law of the survival of the fittest is applied to a population of individuals. This natural optimization method is used to optimize either the response surface approximation equation or the error between the FE model and the measured data. GA is implemented by generating a population and creating a new population by performing the following procedures: (1) crossover; (2) mutation; (3) and reproduction. The details of these procedures can be found in Holland[21] and Goldberg[22]. The crossover operator mixes genetic information in the population by cutting pairs of chromosomes at random points along their length and exchanging over the cut sections. This has a potential of joining successful operators together. Arithmetic crossover technique[22] is used in this paper. Arithmetic crossover takes two parents and performs an interpolation along the line formed by the two parents. For example if two parents *p1* and *p2* undergo crossover, then a random number *a* which lies in the interval [0,1] is generated and the new offsprings formed are *p1(a-1)* and *pa*.

Mutation is a process that introduces to a population, new information. Non-uniform mutation[22] was used and it changes one of the parameters of the parent based on a non-uniform probability distribution. The Gaussian distribution starts with a high variance and narrows to a point distribution as the current generation approaches the maximum generation.

Reproduction takes successful chromosomes and reproduces them in accordance to their fitness functions. In this study normalized geometric selection method was used[22]. This method is a ranking selection function which is based on the normalized geometric distribution. Using this method the least fit members of the population are gradually driven out of the population. The basic genetic algorithm was implemented in this paper as follows:

1) Randomly create an initial population of a certain size.

2) Evaluate all of the individuals in the population using the objective function in equation 4.

3) Use the normalized geometric selection method to select a new population from the old population based on the fitness of the individuals as given by the objective function.

4) Apply some genetic operators, non-uniform mutation and arithmetic crossover, to members of the population to create new solutions.

5) Repeat steps 2-6, which is termed one generation, until a certain fixed number of generations has been achieved

The next section describes simulated annealing which is used to update an FE model using a FE model.

## Simulated Annealing

Simulated Annealing is a Monte Carlo method that is used to investigate the equations of state and frozen states of *n* degrees of freedom system[23]. SA was inspired by the process of annealing where objects, such as metals, re-crystallize or liquids freeze. In the annealing process the object is heated until it is molten, then it is slowly cooled down such that the metal at any given time is approximately in thermodynamic equilibrium. As the temperature of the object is lowered, the system becomes more ordered and approaches a *frozen* state at T=0. If the cooling process is conducted insufficiently or the initial temperature of the object is not sufficiently high, the system may become quenched forming defects or freezing out in metastable states. This indicates that the system is trapped in a local minimum energy state.

The process that is followed to simulate the annealing process was proposed by Metropolis *et al.*[24] and it involves choosing the initial state with energy $E_{old}$ (see equation 4) and temperature $T$ and holding $T$ constant and perturbing the initial configuration and computing $E_{new}$ at the new state. If $E_{new}$ is lower than $E_{old}$, then accept the new state, otherwise if the opposite is the case then accept this state with a probability of *exp -(dE/T)* where *dE* is the change in energy. This process can be mathematically represented as follows:

$$if\ E_{new} < E_{old}\ accept\ state\ E_{new}$$

$$else\ accept\ E_{new}\ with\ probability\ \ exp\left(\frac{E_{new} - E_{old}}{T}\right) \quad (7)$$

This processes is repeated such that the sampling statistics for the current temperature is adequate, and then the temperature is decreased and the process is repeated until a frozen state where *T=0* is achieved.

SA was first applied to optimization problems by Kirkpatrick, *et al.*[23]. The current state is the current updating solution, the energy equation is the objective function in equation 4, and the ground state is the global optimum solution.

## Example: Asymmetrical H-structure

An unsymmetrical H-shaped aluminum structure shown in Figure 3 was used to validate the proposed method. This structure was also used by Marwala and Heyns[4] as well as Marwala[25]. This structure had three thin cuts of 1mm that went half-way through the cross-section of the beam. These cuts were introduced to elements 3, 4 and 5. The structure with these cuts was used so that the initial FE model gives data that are far from the measured data and, thereby test the proposed proce-dure on a difficult FE model updating problem. The structure was suspended using elastic rubber bands. The structure was excited using an electromagnetic shaker and the response was measured using an accelerometer. The structure was divided into 12 elements. It was excited at a position indicated by double-arrows, in Figure 3, and acceleration was measured at 15 positions indicated by single-arrows in Figure 3. The structure was tested freely-suspended, and a set of 15 frequency response functions were calculated. A roving accelerometer was used for the testing. The mass of the accelerometer was found to be negligible compared to the mass of the structure. The number of measured coordinates was 15.

Thereafter, the finite element model was constructed using the Structural Dynamics Toolbox[26]. The FE model used Euler-Bernoulli beam elements. The FE model contained 12 elements. The moduli of elasticity of these elements were used as updating parameters. When the FE updating was implemented the moduli of elasticity was restricted to vary from $6.00 \times 10^{10}$ to $8.00 \times 10^{10}$ N.m$^{-2}$. The weighting factors, in the first term in equation 4, were calculated for each mode as the square of the error between the measured natural frequency and the natural frequency calculated from the initial model and the weighting function for the second term in equation 4 was set to 0.75. When the RSM, SA and GA were implemented for model updating the results shown in Table 1 were obtained.

On implementing the proposed RSM, the FE model was run 150 times to generate the data for functional approximation. The MLP implemented had 12 input variables corresponding to the 12 elements in the FE model, 8 hidden units and one output unit corresponding to the error in equation 4. As described before, the MLP had a hyperbolic tangent activation function in the hidden layer and linear activation function in the output layer. The RSM functional approximation via the MLP was evaluated 10 times (iterations) each time using the GA to calculate the optimum point and evaluating this optimum point on the FE model and then storing the previous optimum point in the data set for the current functional approximation. The scaled conjugate gradient method was used to train the MLP, primarily because of its computational efficiency[27]. The initial functional approximation was obtained by training the MLP for 150 training cycles and on a subsequent functional approximation, where the data set had the previous optimum solution added to it, used 5 training cycles. On using the RSM, the MLP was only initialized once. The GA was implemented on a population size of 50 and 200 generations. The normalized geometric distribution was implemented with a probability of selecting the best candidate set to 8%, mutation rate of 0.3% and crossover rate of 60%.

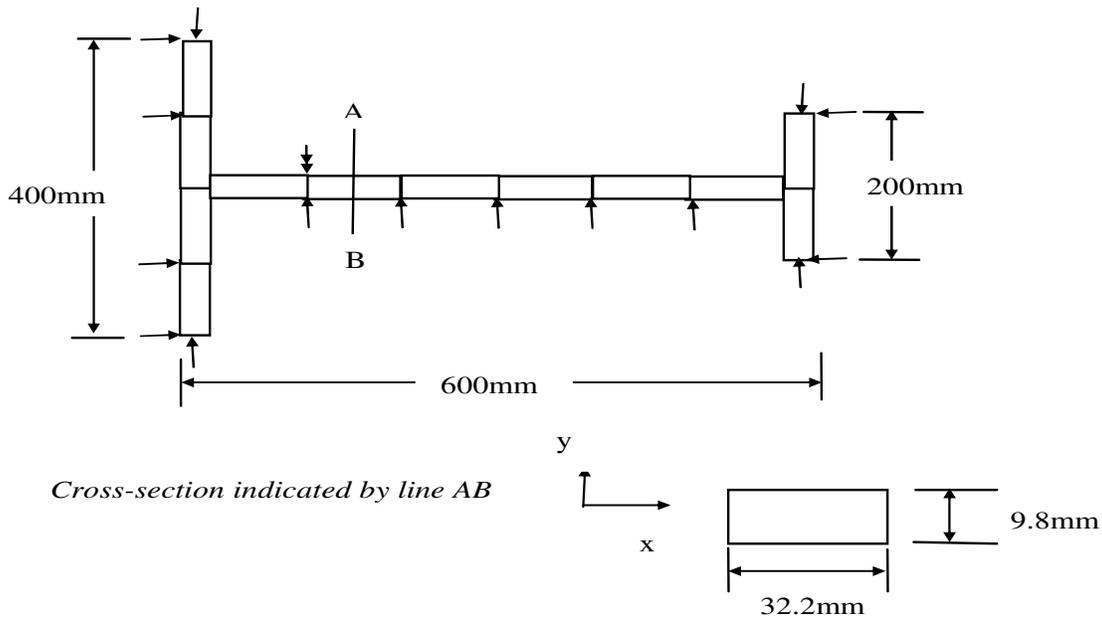

Figure 3. Irregular H-shaped structure

Table 1. Results showing measured frequencies, the initial frequencies and the frequencies obtained when the FE model is updated using the RSM, SA and GA

| Measured Freq (Hz) | Initial Freq (Hz) | Frequencies from RSM Updated Model (Hz) | Frequencies from SA Updated Model (Hz) | Frequencies from GA Updated Model (Hz) |
|---|---|---|---|---|
| 53.9 | 56.2 | 52.2 | 54.0 | 53.9 |
| 117.3 | 127.1 | 118.4 | 118.8 | 120.1 |
| 208.4 | 228.4 | 209.4 | 209.7 | 211.3 |
| 254.0 | 263.4 | 251.1 | 253.8 | 253.4 |
| 445.1 | 452.4 | 432.7 | 435.8 | 438.6 |

When SA and a full FE model was implemented for FE updating, the scale of the cooling schedule was set to 4 and the number of individual annealing runs was set to 3. When the simulation was run, the first run involved 7008 FE model calculations, in the second run 6546 FE model calculations and in the third run 5931 FE model calculations were made.

On implementing the GA and a full FE model, the same options as those that were used in the implementation of the RSM were used. The results showing the moduli of elasticity of the initial FE model, RSM updated FE model, SA updated FE model and GA updated FE model are shown in Figure 4. Table 1 shows the measured natural frequencies, initial natural frequencies and natural frequencies obtained by the RSM, SA and GA updated FE models. The error between the first

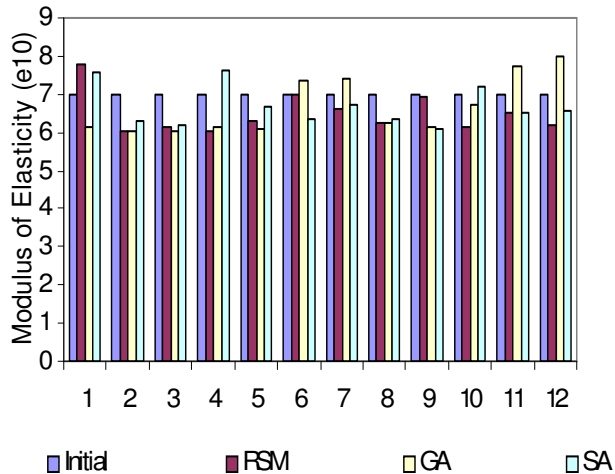

Figure 4. Graph showing the initial moduli of elasticity and the moduli of elasticity obtained when the FE model is updated using the RSM, GA and SA. Here e10 indicates 10 to the power 10 and the units are $Nm^{-2}$

measured natural frequency and that from the initial FE model, which was obtained when the modulus of elasticity of $7.00 \times 10^{10}$ $N.m^{-2}$ was assumed, was 4.3%. When the RSM was used for FE updating, this error was reduced to 3.1% while using SA it was reduced to 0.2% and using the GA approach it was reduced to 0%. The error between the second measured natural frequency and that from the initial model was 8.4%. When the RSM was used, this error was reduced to 0.9% while using SA it was reduced to 1.3% and using the GA it was reduced to 2.4%. The error of the third natural frequencies between the measured data and the initial FE model was 9.6%. When the RSM was used, this error was reduced to 0.5% while using SA reduced it to 0.6% and using the GA and a full FE model reduced it to 1.4%. The error between the fourth measured natural frequency and that from the initial model was 3.7%. When the RSM was used for FE updating, this error was reduced to 1.1% while using the SA reduced it to 0.1% and using the GA and a full FE model reduced it to 0.2%. The error between the fifth measured natural frequency and that from the initial model was 1.6%. When the RSM was used, this error was increased to 2.8% while using SA increased it to 2.1% and using the GA and a full FE model the error was reduced to 1.5%. Overall, the SA gave the best results with an average error, calculated over all the five natural frequencies, of 0.9% followed by the GA with an average error of 1.1% and then RSM with an average error of 1.7%. All the three methods on average improved when compared to the average error between the initial FE model and the measured data, which was 5.5%.

The updated FE models implemented were also validated on the mode shapes they predicted. To make this assessment possible the MAC[11] was used. The mean of the diagonal of the MAC vector was used to compare the mode shapes predicted by the updated and initial FE models to the measured mode shapes. The average MAC calculated between the mode shapes from an initial FE model and the measured mode shapes was 0.8394. When the average MAC was calculated between the measured data and data obtained from the updated FE models, it was observed that the RSM, SA and GA updated FE model gave the improved average of the diagonal of the MAC matrix of 0.8413, 0.8430 and 0.8419, respectively. Therefore, the SA gave the best MAC followed by the GA which was followed by the RSM. However, these differences in accuracies of the MAC and natural frequencies were not significant.

The computational time taken to run the complete RSM method was 46 CPU seconds, while the SA and a full FE model took 19 CPU minutes to run and the GA and a full FE model took 117 CPU seconds. The RSM was found to be faster than the GA which was in turn much faster than the SA which was faster that the GA. On implementing the RSM, 160 FE model evaluations were made, while on implementing the SA 19485 FE model evaluations were made and on implementing the GA 10000 FE model calculations were made. In this paper, a simple FE model with 39 degrees of freedom is updated. It can, therefore, be concluded that if the FE model had several thousand degrees of freedom, the RSM will be substantially faster than the other methods. This conclusion should be understood in the light of the fact that FE models usually have many degrees of freedom.

## Conclusion

In this study, RSM is proposed for FE model updating. The proposed RSM was implemented within the framework of the MLP for functional approximation and GA for optimization of the MLP response surface function. This procedure was compared to the GA and SA. When these techniques were tested on the unsymmetrical H-shaped structure, it was observed that the RSM was faster than the SA and GA without much compromise on the accuracy of the predicted modal properties.

## Acknowledgment

The author would like to thank Stefan Heyns, the now National Research Foundation as well as the AECI, Ltd for their assistance in this work.